%% file: Mixture08.tex
\documentclass[a4paper]{imsart}
\usepackage{natbib}
\usepackage[pdftex,colorlinks]{hyperref}  
\usepackage{amsmath,amssymb,amsthm}
\usepackage{moreverb}
\usepackage{color}

\usepackage[ruled,vlined]{algorithm2e}
\usepackage{rotate}
\usepackage{rotating}
\usepackage{multirow}
\usepackage{graphicx}
\usepackage{fancyheadings}
\usepackage{url}

\newcommand{\argmax}{\mathop{\mathrm{argmax}}}

\def\argmax{\mathop{\mathrm{Argmax}}}

\newcommand{\ER}{Erd\"{o}s-R\'{e}nyi }

\newcommand{\Bcal}{\mathcal{B}}

\newcommand{\Ncal}{\mathcal{N}}

\newcommand{\alphabf}{\mbox{\mathversion{bold}{$\alpha$}}}

\newcommand{\Zbf}{{\bf Z}}
\newcommand{\Xbf}{{\bf X}}

\begin{document}
\begin{frontmatter}

\title{Clustering based on Random Graph Model embedding Vertex Features}
 \begin{aug}
   \author{\fnms{Hugo} \snm{Zanghi}},
    \author{\fnms{Stevenn} \snm{Volant}} and 
    \author{\fnms{Christophe} \snm{Ambroise}}

    \ead[label=e1]{hugo.zanghi@exalead.com}
    \ead[label=e2]{stevenn.volant@agroparistech.fr}
    \ead[label=e3]{christophe.ambroise@genopole.cnrs.fr}
    \ead[label=u]{http://www.exlead.com}

    \address{Exalead\\
      10, place de la Madeleine\\
      75008 \'Paris, FRANCE\\
     \printead{u}\\
     \printead{e1,e2,e3}}

    \runauthor{H. Zanghi, S. Volant and C. Ambroise}
    \runtitle{Clustering using structure and features}


\end{aug}

\input{abstract}

\end{frontmatter}

\input{introduction}

\input{model_and_likelihoods}

\input{experiments}

\input{conclusion}

\newpage
\bibliographystyle{harvard}
\bibliography{Mixture08}

\end{document}

%% file: abstract.tex
\begin{abstract}
Large datasets with  interactions between objects  are common to
numerous scientific fields (i.e. social science, internet, biology\ldots).
The interactions naturally define a graph and a common way to explore or summarize such dataset
is graph clustering.
Most   techniques for clustering graph vertices 
just use  the topology of connections ignoring informations in the vertices features.
In this paper,  we provide a clustering algorithm exploiting both types of data based on a 
statistical model with latent structure  characterizing each vertex both by a vector of features
as well as by its connectivity. We perform simulations to compare our algorithm with existing approaches, 
and also evaluate our method with real datasets based on hyper-textual documents. We find that our algorithm 
successfully exploits whatever information is found both in the connectivity pattern and in the features.
\end{abstract}

%% file: introduction.tex
\section{Introduction}


Classical data analysis has been developed for sets of objects with 
features, but when  explicit relationships exist between objects, classical 
data analysis cannot take these  relations into account. On the other hand, 
much recent research has been performed for  analyzing graphs, for example in 
finding relationships in social sciences, gene interactions in biology and hyperlinks 
analysis in computer science, providing insights into the interactions  in these 
networks. Many approaches to graph analysis have been proposed. Model-based approaches, 
\textit{i.e.}, methods which rely on a statistical model of network edges and 
vertices, such as those first proposed by \ER, often allow to get insight into the network 
structure deducing their internal properties.

An interesting alternative to the basic \ER model which does not fit well to
real networks is to consider a mixture of distributions
\citep{Harary1982,Snijders1997,Newman2007, Daudin2008} where it is
assumed that nodes are spread among an unknown number of latent connectivity
classes. Conditional on the hidden class label, edges are still independent and
Bernoulli distributed, but their marginal distribution is a mixture
of Bernoulli distributions with strong dependence between the
edges. Many names have been proposed for this model, and in the
following, it will be denoted by MixNet, which is equivalent to
Block Clustering of \citet{Snijders1997}. Block-Clustering for
classical binary data can be dated back to  early work in the seventies
\citep{White1971, Govaert1977}.

But vertex content is also sometimes available in addition to the network information used in 
the methods mentioned above. A typical example is the world-wide-web which can be
described by either hyperlinks between web pages or by the words
occurring in the web pages: each vertex represents a web page
containing the occurrences of some words and each directed edge a
hyperlink.  The additional information represented by the vertex features is rarely used in
network clustering but can provide crucial information. Here we combine information from both vertex 
content traditionally used in classical data analysis to information found in the graph structure, in 
order to cluster objects into coherent groups. This paper proposes a statistical model, called 
\texttt{CohsMix} (for \emph{\textbf{Co}variates on \textbf{h}idden \textbf{s}tructure using \textbf{Mix}ture models}),
which considers the dependent nature of the data and the relation with vertex features (or covariates) in order to capture 
a hidden structure.

Considering spatial or relational data neighbourhood is not an
original approach in clustering. For instance,  Hidden
Markov Random Fields (HMRF)  are  well adapted to handle spatial
data and are  widely used in image analysis. When the spatial network is
not given it is  generally obtained using Delaunay triangulation
\citep{ambroise1997csd}.

\citet{hoff2003rem} proposed a new way to deal with covariates. He
suggested to model the expected value of the relational ties by a
logistic regression. The problem of this method is the dependence
between the observations conditional on the regression parameters and
the covariates. Hence, he proposed to incorporate random effect
structures in a generalized linear model setting. The distribution of
dependence among the random effects determines the dependence among
the edges.

There are also approaches based on non statistical frameworks. In
particular, it is noticeable that there exits a strong similitude
between multiple view and graph models with covariates. In fact,
multiple view learning algorithms \citep{ruping2005lmv} consider
instances which have multiple representations and simultaneously
exploit these views to find a consensus partition.

The second section introduces the proposed model, which is an extension
to the MixNet model. Since the model considers a great number of
dependencies, the proposed estimation scheme proposes a variational
approach of the EM algorithm. This approach allows us to deal with
larger network than the Bayesian framework.  Then we introduce
practical strategies for the initialization and the choice of the
number of groups. In the third section extensive simulations
illustrate the efficiency of the this algorithm and real datasets
dealing with hyper-textual documents are studied.
A R package named \texttt{CohsMix} is available upon request.

%% file: model_and_likelihoods.tex
\section{A Mixture of Network with covariates}
This section introduces the proposed model. We choose to
consider a model which assumes independence of covariates
and edges conditional on the node classes. It assumes that both the
connectivity pattern and the vertex features can be  explained by the class.
In the web context this model considers that  a given class contains documents  which
have both a similarity between   occurring words and a similarity of connectivity pattern
with documents inside and outside the class. Although this assumption
does not explicitly  model  the idea that authors tend to link similar topics
(occurring words) which creates a thematic locality \citep{davison2000},
it allows us to detect clusters of local theme. Its simplicity makes it a robust well
adapted model  to  the real web.

\subsection{Models and Notation}
Let us define a random graph $G$, where $\mathcal{V}$ denotes the set
of  vertices.  Based on the MixNet model, our model assumes that
$\mathcal{V}$  is  partitioned into $Q$ hidden classes.
Let us  denote by $Z_{iq}$ the indicator  variable such that $\{Z_{iq}=1\}$ if node $i$ belongs to class $q$.
$\Zbf=(\Zbf_1,\hdots,\Zbf_n)$ is the vector of random independent indicator  variables such that
\begin{equation}
  \Zbf_i \sim {\cal M}(1,\alphabf=\{\alpha_1,...,\alpha_Q\}),\label{eq:distOfZ}
\end{equation}
with $\alphabf$ the vector of class  proportions.
Edges
are Bernoulli random variables
\begin{equation}
 X_{ij}|{Z_{iq} Z_{jl}=1} \sim \Bcal(\pi_{ql})\label{eq:distOfX},
\end{equation}
conditionally independent, given the
 node classes
$$
P (\Xbf|\Zbf ) = \prod_{ij} \prod_{q,l}  P (X_{ij} | Z_{iq}Z_{jl}=1 ) ^{Z_{iq}Z_{jl}}.
$$


In this paper, we consider an undirected graph we suppose that there is no 
self-loops, i.e. a node can not be connected to itself ($X_{ii}$ = 0). Nevertheless, the
method can easily be generalized to encompass directed graphs  with
self-loops.

\paragraph{Vertex Features.}
Hereafter we consider $n$ objects described both by their connections
and $p$ features. In that case the data under study can be
represented into different forms. One might for example consider a
two part vector for characterizing each object, where  the first part
contains the feature of the object $\mathbf{Y}_i$  and the second
part contains a binary vector representing the connection to all $n-1$
other objects $\mathbf{X}_i$. Continuing our example about world-wide-web,
the web pages can be viewed as a vector of word occurrences with hyperlinks
or as two matrices. One based on the adjacency matrix describing the
topology of the graph generated by the hyperlinks and the other by
the features matrix generated by the word occurrences in each web page.

Hereafter we consider that the  $p$ dimensional feature vector  associated to object $i$
is defined by :
$$
\mathbf{Y}_i = \left (
   \begin{array}{cccc}
      Y^{(1)}_i  \\
      Y^{(2)}_i \\
      \vdots \\
      Y^{(p)}_i \\
   \end{array}
   \right )
$$

We assume that the feature vectors $\mathbf{Y}_i$ are multivariate normally distributed
\begin{equation}
 \mathbf{Y}_i|Z_{iq}=1 \sim \Ncal(\boldmath{\mu}_q, \Sigma_q  )\label{eq:distOfY}
\end{equation}
where

$
\boldmath{\mu}_q = \left (
   \begin{array}{cccc}
      \mu^{(1)}_q  \\
      \mu^{(2)}_q \\
      \vdots \\
      \mu^{(p)}_q \\
   \end{array}
   \right )
$
and
$
\mathbf{\Sigma_q} =\sigma I
$
the covariance matrix is proportional to the identity.

The  random feature vectors  $\mathbf{Y}_i$ are 
 conditionally independent, given the node classes
$$
 P(\mathbf{Y} | \mathbf{Z}) = \prod_i \prod_q P(\textbf{Y}_i|Z_{iq})^{Z_{iq}}.
$$
The conditional distribution associated to covariates can be written as follow :
\begin{eqnarray*}
\label{distY}
\log P(\mathbf{Y} | \mathbf{Z})& = &\sum_i \sum_q Z_{iq}\log P(\mathbf{Y}_i|Z_{iq})\\
    & = & \sum_i \sum_q Z_{iq} \left[ \left(\log\frac{1}{2\pi^{\frac{n}{2}} det(\Sigma)^{\frac{1}{2}}}\right) - \frac{1}{2} (\textbf{Y}_i - \boldmath{\mu}_q)^T \sigma^{-1} (\textbf{Y}_i - \boldmath{\mu}_q) \right].
\end{eqnarray*}

\begin{figure}
  \centering
\includegraphics[width=5cm]{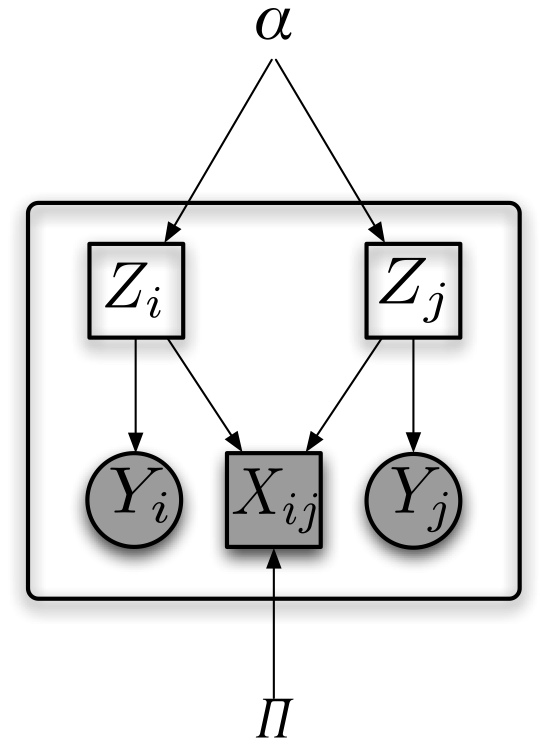}
  \caption{Graphical representation of the \texttt{CohsMix} Model. The squares
      represent discrete random variables and circles continuous random
     variables.
     \label{fig:model3}}
\end{figure}

The proposed mixture model assumes independence of $\mathbf{X}$ and
$\mathbf{Y}$ conditional on $\mathbf{Z}$. Considering this independence
between edges and covariates, the complete log-likelihood can be
written as (Figure \ref{fig:model3}):
\begin{eqnarray*}
P(\mathbf{X},\mathbf{Y},\mathbf{Z})  = P(Z) P(\mathbf{X},\mathbf{Y}|Z)  
& = &  P(\mathbf{Z}) P(\mathbf{Y}|\mathbf{Z}) P(\mathbf{X}|\mathbf{Z}).
\end{eqnarray*}
The next section proposes an estimation scheme for the CohsMix model.

\subsection{Variational \textsc{EM} algorithm for CohsMix}
\label{Variational}
In the classical \textsc{em} framework developed by \citet{dempster1977},
where  $\mathbf{X}$ and $\mathbf{Y}$ are the available data, the inference of
the unknown  parameters $\mathbf{\Theta}$ spread over a latent structure  
$\mathbf{Z}$ uses   the  following conditional expectation:
\begin{multline}
  \label{eq:conditional_expectation}
  Q\left(\mathbf{\Theta}|\mathbf{\Theta}^{(m)}\right)                           =
  \mathbb{E}        \left\{       \log
    \mathcal{L}_c(\mathbf{X},\mathbf{Y},\mathbf{Z}; \mathbf{\Theta}) \big| \mathbf{X}, \mathbf{Y};
    \mathbf{\Theta}^{(m)}\right\} \\
  =                                    \sum_{\mathbf{Z}\in\mathcal{Z}}
  \mathbb{P}\left(\mathbf{Z} \big|\mathbf{X},\mathbf{Y}; \mathbf{\Theta}^{(m)}\right) \log
  \mathcal{L}_c(\mathbf{X},\mathbf{Y},\mathbf{Z};\mathbf{\Theta})
\end{multline}

where

$$
\mathbf{\Theta^{(m+1)}} = \displaystyle{\argmax_{\mathbf{\Theta}}} Q(\mathbf{\Theta}, \mathbf{\Theta}^{(m)}).
$$

The  usual \textsc{em} strategy would  be to alternate an \textsc{E}-step
computing the conditional expectation \eqref{eq:conditional_expectation}  
with  an \textsc{M}-step  maximizing this quantity over the parameter of  
interest $\mathbf{\Theta}$.
Unfortunately, no closed form of $Q\left(\mathbf{\Theta}|\mathbf{\Theta}^{(m)}\right)$  
can be formulated in the present case. The technical difficulty lies in the
complex dependency structure of the model. Indeed, $\mathbb{P}
(\mathbf{Z}|\mathbf{X},\mathbf{Y}; \mathbf{\Theta})$  cannot be factorized, as argued in \citet{Daudin2008}.
This makes the direct calculation of $Q\left(\mathbf{\Theta}|\mathbf{\Theta}^{(m)}\right)$  
impossible. To tackle this problem we use a variational approach \citep[see, e.g.,]
[for elementary results on variational methods]{jordan1999}. In this framework,  
the conditional distribution of the latent variables
$\mathbb{P} (\mathbf{Z}| \mathbf{X},\mathbf{Y}; \mathbf{\Theta}^{(m)})$  is approximated by a more
convenient  distribution  denoted  by  $R(\mathbf{Z})$,  which  is
chosen carefully in order to be tractable. Hence, our \textsc{em}-like
algorithm  deals with the  following approximation of the conditional expectation
\eqref{eq:conditional_expectation}
\begin{equation}
  \label{eq:conditional_expectation_approx}
  \mathbb{E}_{R}\left\{\log\mathcal{L}(\mathbf{X},\mathbf{Y},\mathbf{Z};\mathbf{\Theta})
  \right\}  =  \sum_{\mathbf{Z}\in\mathcal{Z}}  R(\mathbf{Z} )  \log
  \mathcal{L}(\mathbf{X},\mathbf{Y}, \mathbf{Z}; \mathbf{\Theta}).
\end{equation}

In the following section we develop a variational  argument in order
to choose an approximation $R(\mathbf{Z})$ of
$\mathbb{P}(\mathbf{Z}|\mathbf{X},\mathbf{Y};\mathbf{\Theta}^{(m)})$. This enables us to compute the
conditional  expectation \eqref{eq:conditional_expectation_approx} and
proceed to the maximization step.

\subsection[Variational estimation of the latent structure]
{Variational estimation of the latent structure(\textsc{E}-step)}\label{sec:Estep}

In this part, $\mathbf{\Theta}$ is assumed  to be known, and we are looking
for an  approximate distribution  $R(\cdot)$ of the  latent variables.
The variational approach consists in maximizing a lower bound
$\mathcal{J}$  of the  log-likelihood $\log \mathbb{P}(\mathbf{X},\mathbf{Y}; \mathbf{\Theta})$,
defined as follows:
\begin{equation}
  \label{eq:definition_J}
  \mathcal{J}\left( \mathbf{\Theta}\right)    =
  \log\mathbb{P}(\mathbf{X},\mathbf{Y}; \mathbf{\Theta})
  -   \mathrm{D}_{KL}\left\{R(\mathbf{Z})  \|   \mathbb{P}(\mathbf{Z}  | \mathbf{X},\mathbf{Y};
    \mathbf{\Theta}^{(m)})\right\}
\end{equation}
where $\mathrm{D}_{KL}$ is the K\"ullback-Leibler divergence. This
measures the  difference between the probability distribution
$\mathbb{P}(\cdot|\mathbf{\Theta})$ in the underlying model and its
approximation $R(\cdot)$. An intuitively straightforward choice for $R(\cdot)$
is a completely factorized distribution \citep[see][]{Robin2007,Zanghi2008}
\begin{equation}
  \label{eq:R_factorized}
  R(\mathbf{Z}) = \prod_{i\in \mathcal{P}}  h_{\boldsymbol{\tau}_i}(\mathbf{Z}_i),
\end{equation}
where  $h_{\boldsymbol  \tau_i}$ is  the  density  of the  multinomial
probability    distribution   $\mathcal{M}(1;\boldsymbol\tau_i)$,   and
$\boldsymbol{\tau}_i  =  (\tau_{i1},\dots,\tau_  {iQ})$  is  a  random
vector containing  the variational  parameters to optimize.   The complete
set        of       parameters        $\boldsymbol        \tau       =
\left\{\tau_{iq}\right\}_{i\in\mathcal{P},q\in\mathcal{Q}}$ is what we are seeking
to obtain via the variational  inference. In the  case in
hand the  variational approach intuitively operates  as follows: each
$\tau_{iq}$ can  be seen as  an approximation of the  probability that
vertex $i$ belongs to cluster $q$, conditional on the data, that is,
$\tau_{iq}$ estimates $\mathbb{P}(Z_{iq} = 1 | \mathbf{X},\mathbf{Y};\mathbf{\Theta})$, under the
constraint   $\sum_{q}  \tau_{iq}=1$. In the ideal case where
$\mathbb{P}(\mathbf{Z}|\mathbf{X},\mathbf{Y}; \mathbf{\Theta})$ can be factorized as
$\prod_i\mathbb{P}(\mathbf{Z}_i|\mathbf{X},\mathbf{Y};\mathbf{\Theta})$ and the parameters
$\tau_{iq}$  are  chosen  as  $\tau_{iq}  = \mathbb{P}(Z_{iq}  =  1  |\mathbf{X},\mathbf{Y};
\mathbf{\Theta})$, the K\"ullback-Leibler
divergence is null and the bound $\mathcal{J}$ reaches the log-likelihood.

The lower bound
$\mathcal{J}$ to be maximized in order to estimate $\boldsymbol \tau$ can be expressed as
\begin{eqnarray*}
\mathcal{J}_{\boldsymbol\tau} = \mathbb{E}_{R(Z)} \left\{ \mathcal{J}(\mathbf{\Theta}) \right\} = \mathbb{E}_{R(Z)}\{\log(P(\mathbf{X},\mathbf{Y}, \mathbf{Z}))|\mathbf{X},\mathbf{Y};\mathbf{\Theta}\} - \displaystyle{\sum_Z} R(\mathbf{Z})\log(R(\mathbf{Z})).
\end{eqnarray*}
The  optimal approximate  distribution  $R$ is  then
derived  by  direct  maximization of  $\mathcal{J}_{\boldsymbol\tau}$.
Let all the parameters $\hat{\pi}_{ql}, \hat{\alpha}_q, \hat{\boldmath{\mu}}_{q}$ and $\hat{\sigma}$ be known.
The  following fixed-point relationship  holds  for the optimal  variational  parameters  $\widehat{\boldsymbol\tau} =  \arg
 \max_{\boldsymbol{\tau}} \mathcal{J}_{\boldsymbol\tau}$.
\begin{equation}
\label{eq:tau}
\hat{\tau}_{iq}^{(m+1)}  \propto \hat{\alpha}_q \displaystyle{\prod_{j \neq
    i}         \prod_{l}} \left[        \hat{\pi}_{ql}^{x_{ij}}
  (1-\hat{\pi}_{ql})^{1-x_{ij}} \right]^ {\tau_{jl}^{(m)}} \prod_{k=1}^{p} \left[ \exp(\frac{1}{2\hat{\sigma}^2} \left(-(Y_i^{(k)} - \hat{\boldmath{\mu}}_{q}^{(k)})\right)^2 )\right].
\end{equation}
Once again, the maximization of $\mathcal{J}_{\boldsymbol\tau}$  provides the optimal values of the parameters.
\label{ParameterEstimator}
The optimal parameters $\alpha_q, \pi_{ql}, \boldmath{\mu}_q $ and $\sigma$, i.e. the parameters maximizing $\mathcal{J}_{\boldsymbol\tau}$ satisfy the following relations:

\begin{equation}
\label{eq:estimation}
\begin{aligned}
 \hat{\alpha}_q &= \frac{1}{n} \displaystyle{\sum_{i=1}^n} \tau_{iq},\\
 \hat{\pi}_{ql} &= \frac{\displaystyle{\sum_ {i \neq j}}  \tau_{iq} \tau_{jl} x_{ij}}{\displaystyle{\sum_ {i \neq j}}  \tau_{iq} \tau_{jl}},\\
 \hat{\mu}_q &= \frac{\displaystyle{\sum_i} \tau_{iq} \mathbf{Y}_i}{\displaystyle{\sum_{i}} \tau_{iq}} \ \  \mbox{and} \ \  \hat{\sigma} &= \frac{\displaystyle{\sum_i} \displaystyle{\sum_q} \tau_{iq} (\mathbf{Y}_i - \hat{\zeta}_q)^T (\mathbf{Y}_i - \hat{\zeta}_q)} {\displaystyle{\sum_{i}} \displaystyle{\sum_{q}} \tau_{iq}}.
\end{aligned}
  \end{equation}
For completeness, we summarize the variational \textsc{EM} algorithm for \texttt {CohsMix} in the Algorithm \ref{algo:varEm}.
 \begin{algorithm}\label{algo:varEm}
 \dontprintsemicolon
 \KwData{Matrices of connectivities $\mathbf{X}$ and similarities $\mathbf{Y}$}

  \CommentSty{/* Initialization of the parameters */}\;
  \BlankLine
  $\mathbf{\Theta}^{(0)} = \left(\alpha_{1}^{(0)},...,\alpha_{Q}^{(0)}, \pi_{11}^{(0)},...,\pi_{QQ}^{(0)},
  \boldmath{\mu}_{1}^{(0)},...,\mu_{p}^{(0)},\sigma^{(0)} \right), m=0$\;
  \BlankLine
  \While{not convergence}{
    \CommentSty{/* \textsc{E}stimation step */}\;
    \CommentSty{/* Compute $\boldsymbol        \tau       = \left\{\tau_{iq}\right\}_{i\in\mathcal{P},q\in\mathcal{Q}}$
     the probabilities that vertex $i$ belong to cluster $q$ finding fix point of $g()$ */}\;
    \ForEach{$i \in \{1,...,N\}$}{
        \ForEach{$q \in \{1,...,Q\}$}{
        $\tau_{iq}^{(m+1)} = g(\boldsymbol{\tau}^{(m)})$ (see Equation \ref{eq:tau})\;
        }
        \CommentSty{/* normalize posterior probabilities */}\;
    $scale = \sum_{q=1}^Q \tau_{iq}$ \;
    $\tau_{iq} = \tau_{iq}  \frac{1}{scale}$, $\forall q \in \{1,...,Q\}$ \;
    }
    \CommentSty{/* \textsc{M}aximization step */}\;
    \CommentSty{/* re-estimate the distribution parameters to maximize the likelihood of the data */}\;
       Update parameters according Equation \ref{eq:estimation} :\;
    \ForEach{$q \in \{1,...,Q\}$}{
              $\alpha_{q} ^{(m+1)}= \argmax_{\alpha_{q}}  \mathcal{J}_{\boldsymbol\tau} (\mathbf{\Theta})$ \;
        \ForEach{$l \in \{1,...,Q\}$}{
       $\pi_{ql} ^{(m+1)}= \argmax_{\pi_{ql}}  \mathcal{J}_{\boldsymbol\tau} (\mathbf{\Theta})$ \;
       $\mu_{q}^{(m+1)}= \argmax_{\boldsymbol{\mu_{q}}} \mathcal{J}_{\boldsymbol\tau} (\mathbf{\Theta})$ \;
   }
}    
    m = m +1\;
    \BlankLine
  }
  \BlankLine
  \BlankLine
 \KwResult{Estimated parameters $\mathbf{\Theta}$ and posterior probabilities $\tau_{iq}$}
 \BlankLine
 \caption{Variational \textsc{EM} CohsMix Algorithm}
\end{algorithm}

\subsection{Model selection: ICL algorithm}
As the number of clusters is an unknown parameter of our statistical model, it is possible to use the Integrated
Classification Likelihood (ICL) to choose the optimal number of classes \citep{biernacki2000}. The ICL criterion is
essentially derived from the ordinary BIC considering the complete log-likelihood instead of the log-likelihood.
This optimal number is obtained by running our  algorithm concurrently for models from 2 to Q classes and
selecting the solution which maximizes the ICL criterion.  In our situation where additional covariates are considered,
the ICL criterion can be written as:
\begin{eqnarray*}
ICL(Q) & = & \max_{\mathbf{\Theta}} \log\mathcal{L}(\mathbf{X},\mathbf{Y},\mathbf{Z};\mathbf{\Theta},Q) -\underbrace{\frac{1}{2}\times Q(Q-1)\log(\frac{n(n-1)}{2})}_{\mbox{\footnotesize{related to $\pi_{ql}$}}} -\underbrace{\frac{Q-1}{2}\log(n)}_{\mbox{\footnotesize{related to $\alpha_q$}}} \\
& & -  \underbrace{ p(p-1)\log(\frac{n(n-1)}{2}) + p \times Q\log(\frac{n(n-1)}{2})}_{\mbox{\footnotesize{related to $\boldmath{\mu}_{q}$ \mbox{and} $\sigma$}}}
\end{eqnarray*}
This expression of the ICL criterion is based on the method described in \citet{Daudin2008}.


%% file: experiments.tex
\section{Experiments}
In this section, we report experiments in order to assess the performances and limitations of the proposed model
in a clustering context.  We consider synthetic data generated according to the assumed random graph model,
 as well as real data from  the  web. Using synthetic graphs allows us to evaluate the quality of the
parameter estimation. In parallel, we also compare classification
results with  alternative clustering methods using a ground truth. The real
datasets consist of hypertext documents retrieved from  a  websearch query.
A R package named  \texttt{CohsMix} is available upon request.

\subsection{Comparison of algorithms}
\paragraph{Simulations set-up}
In these experiments, we consider simple affiliation models with two
parameters defining the probability of connection between nodes of the
same class and of different classes, respectively $\pi_{qq}=\lambda$
and $\pi_{ql}=\epsilon$ and equal  mixture proportion $\alpha_1=...=\alpha_Q=\frac{1}{Q}$.
We consider models with $n=150$ nodes.

We generate graph models in order to evaluate the algorithm
performances as the difficulty of the problem varies.  The
clustering problem increases in difficulty with the number of classes
$Q$, the number of features $nbCov$, $d(\lambda, \epsilon)$ the
euclidean distance between intra and extra connectivity parameters and
$d(\mu_q,\mu_l)$ the distance between the feature mean vectors of classes.  We
decide to focus on these parameters to produce data with different
levels of structure and eventually  consider 43 different
graph models whose description are summarized in Table
\ref{table:models}. Each model is simulated 20 times.

\begin{table}
 \centering
 \begin{tabular}{c c c c c}
\hline
Experiments    & $Q$            & $nbCov$      &  $d(\lambda, \epsilon)$  & $d(\mu_q^{(j)},\mu_l^{(j)})$       \\ \hline
       a        &  $\{2,...,12\}$    &   3          &         0.4              &         4              \\
       b        &  5             &   $\{2,...,15\}$ &         0.2              &         4              \\
       c        &  3             &   3          &         $\{0,...,0.5\}$      &          4              \\
       d        &  3             &   3          &         0                &        $\{4,...,8.5\}$     \\ \hline
\end{tabular}
\caption{ \label{table:models} Parameters of the four  different settings  which are used to generate the 43
affiliation models considered in the experiments.}
\end{table}

We use the adjusted Rand Index \citep{huar85} to evaluate the agreement between the estimated and the actual partition.
The Rand index is based on a ratio between the number of node pairs belonging to the same and to different  
classes when considering  both  partitions. It lies between 0 and 1, two identical
 partitions having an adjusted Rand Index equal to 1.

To avoid initialization issues, the algorithm is started with multiple initialization points and the best
result is selected based on its likelihood. Thus, for each simulated graph, the algorithm is run 10 times and
the number of clusters is chosen using the Integrated Classification Likelihood criterion, as proposed in the
previous section.

\paragraph{Alternative clustering methods} Additionally to the \texttt{CohsMix} algorithm study, we compared it with two "rivals" : a
multiple view learning algorithm \citep{ruping2005lmv,zhang2006lpm},
and a Hidden Markov Random Fields \citep{ambroise1997csd}:
\begin{itemize}
\item \textit{Spectral Multiple View Learning (SMVL) :} There exits a strong similitude between multiple view and graph models with covariates. In fact, multiple view learning algorithms consider instances which have multiple  representations and simultaneously exploit these views to find a consensus partition.  This is achieved via spectral clustering on a  linear combination of  a standard kernel  corresponding to the graph structure and a kernel corresponding to    vertex proximity.
\item \textit{Hidden Markov Random Fields (HMRF) :} 
  Hidden Markov Random Fields are commonly used to  handle spatial
  data and are  widely used in image analysis.  We use a classical Potts
  model on the latent structure which encourage spatial smoothing of
  the cluster. This kind of approach uses the graph structure to  smooth the partition of the vertex over the graph, whereas the   approach proposed in this paper uses the graph structure directly to  estimate the vertex partition.
\end{itemize}

\paragraph{Simulations results}

We focus our attention on the Rand Index for each algorithm. Indeed,
a well  estimated partition  leads  to good estimates.

\begin{figure}[htb]
 \centering
 \begin{tabular}{c c}
   \includegraphics[width=6cm]{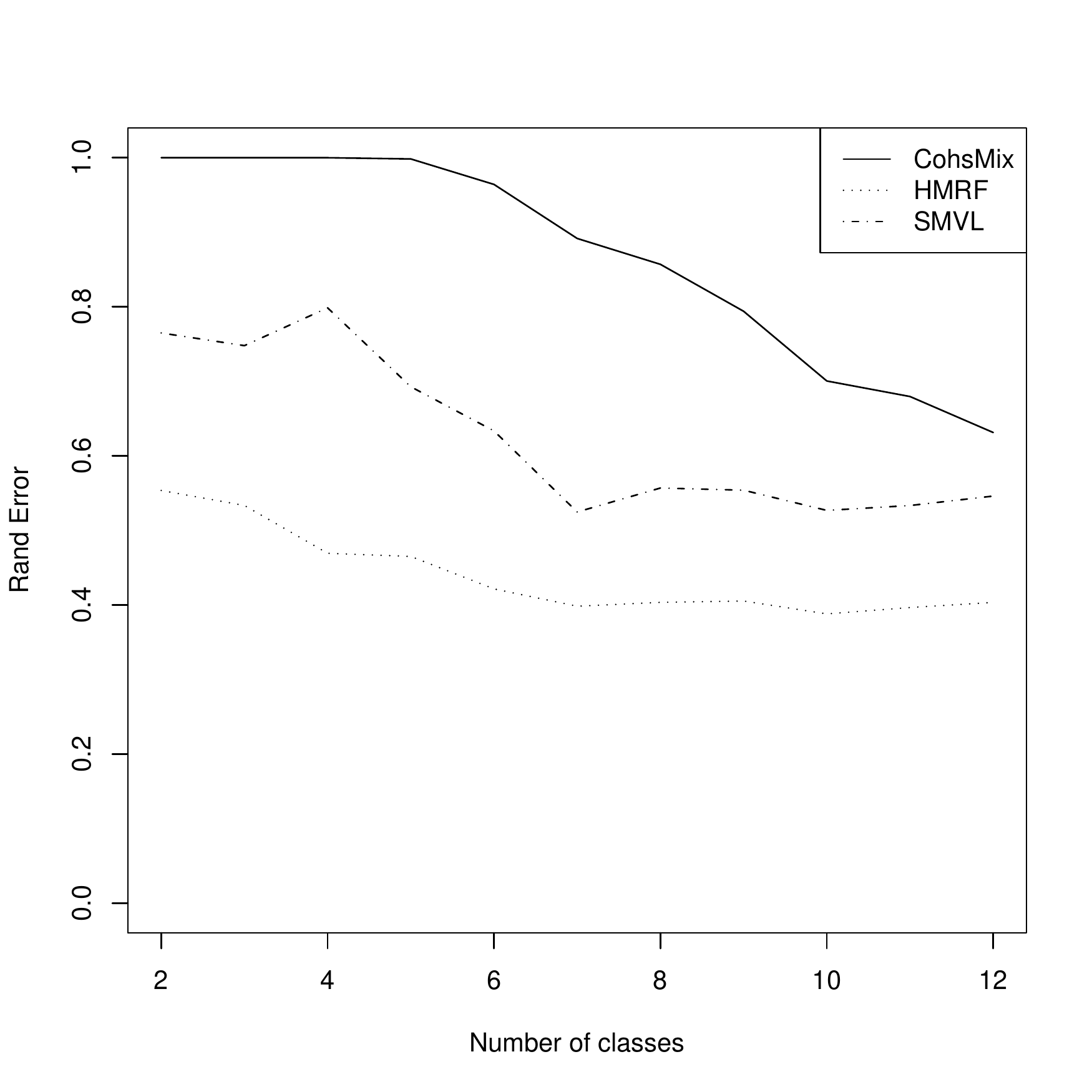} &  \includegraphics[width=6cm]{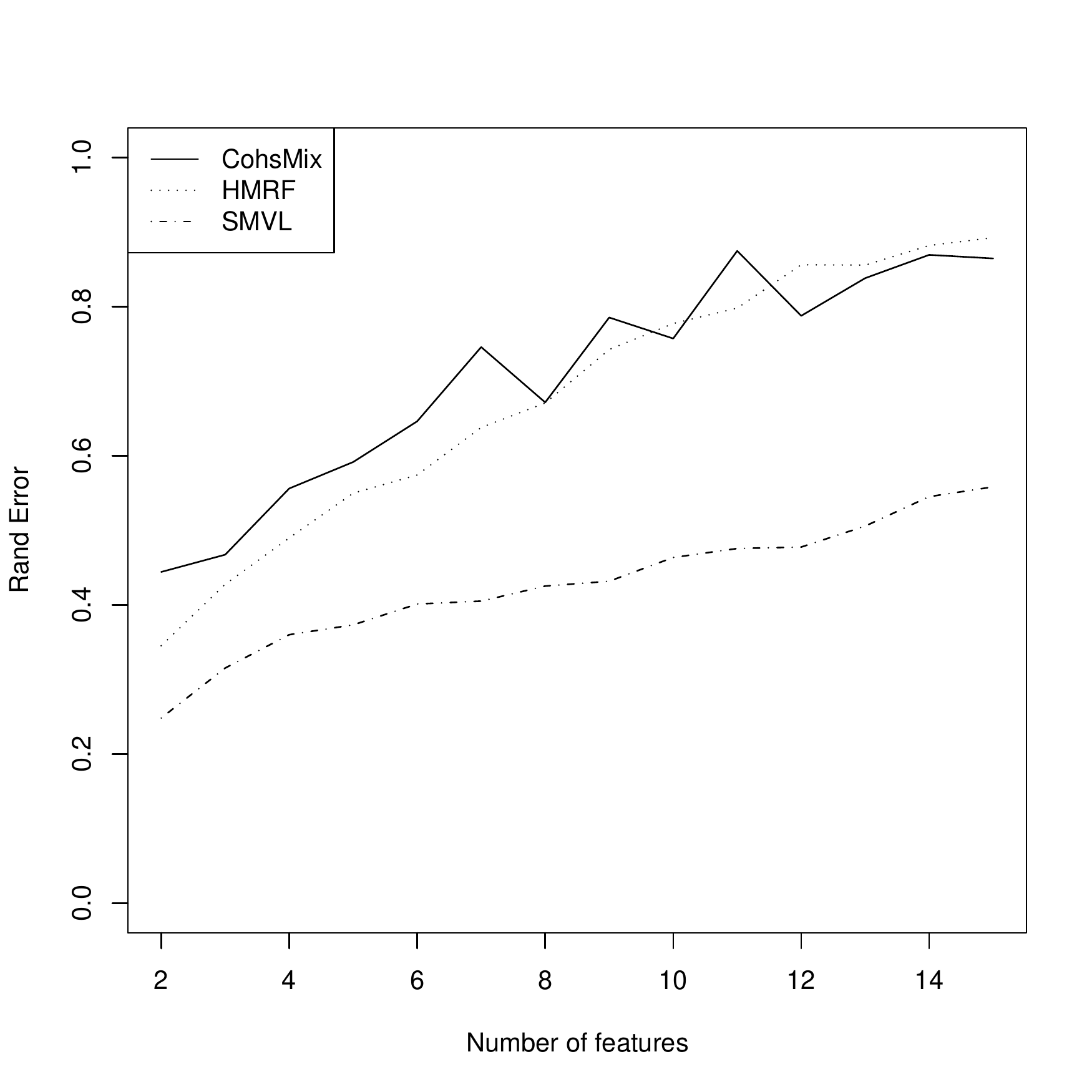} \\
  (a)   & (b) \\
\includegraphics[width=6cm]{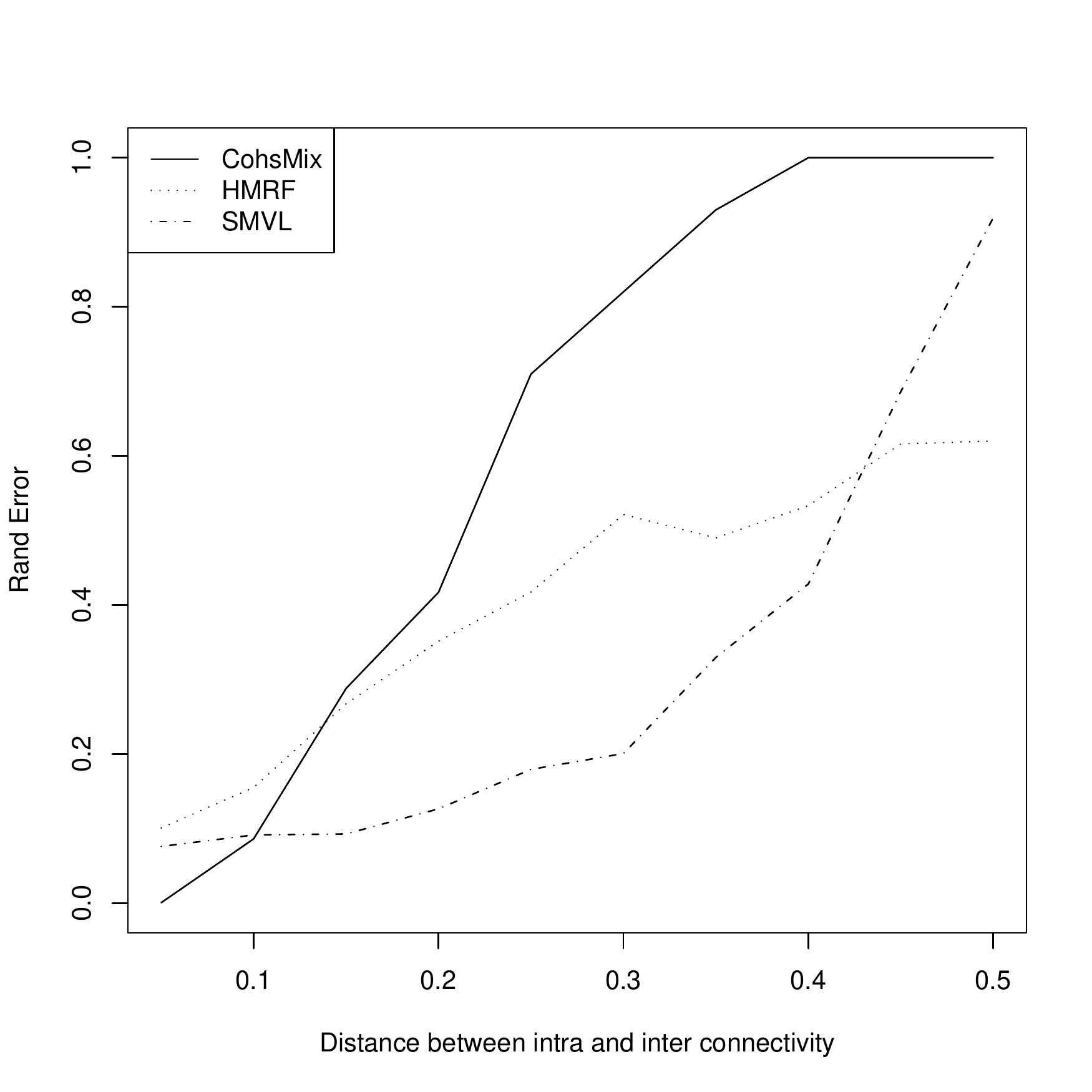}  &\includegraphics[width=6cm]{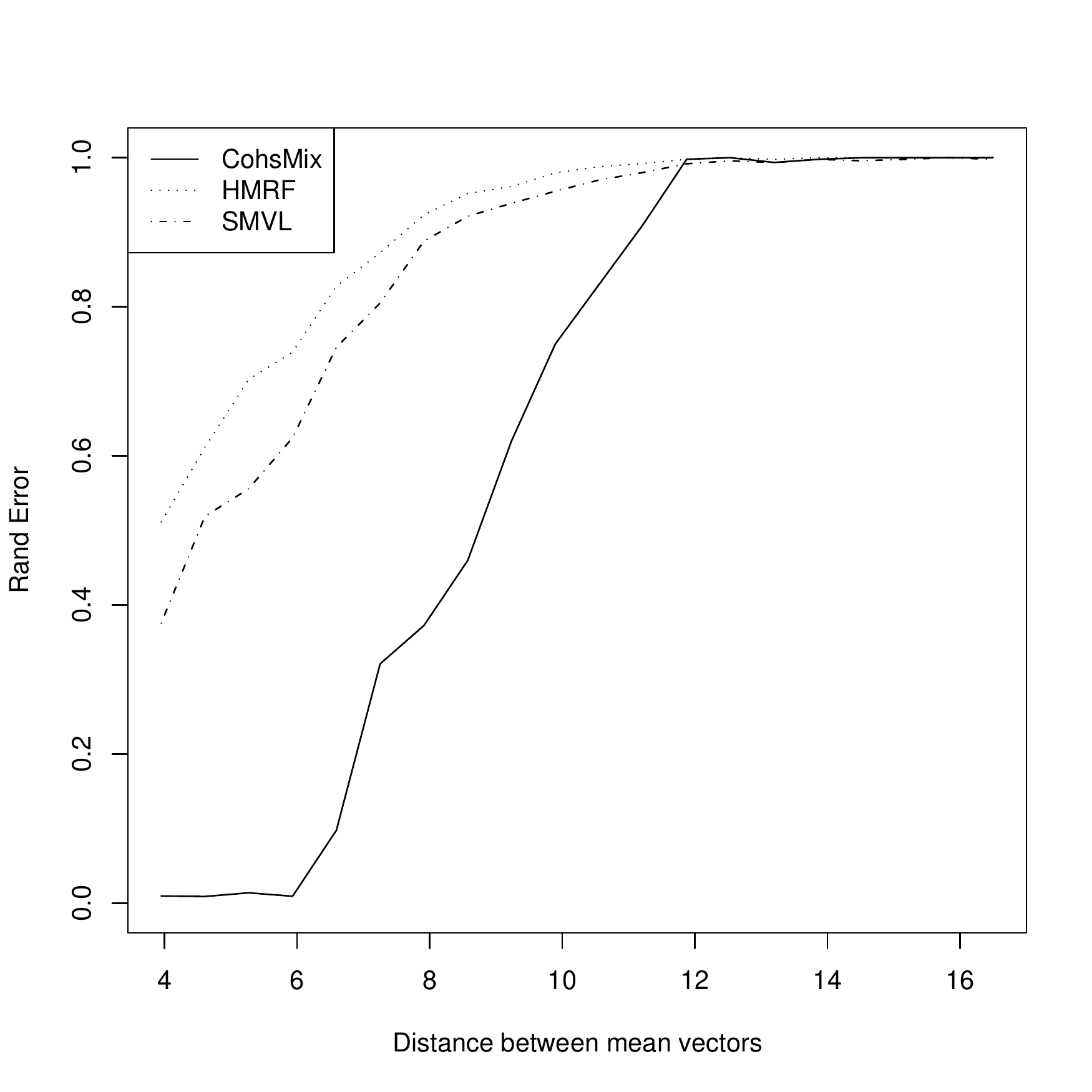} \\
  (c)   & (d) \\
\end{tabular}
 \caption{Comparison of HMRF, Spectral MLV and \texttt{CohsMix}. (a) Varying $Q$ the number of classes. (b) Varying the number of Features. (c) Varying the distance between intra and inter connectivity parameters. (d) Varying the distance between the mean vector of the classes.}
\label{fig:Comp}

\end{figure}

As expected, the performance of the three algorithms
decrease with the number of groups (Figure \ref{fig:Comp} a).

 A first
interesting result is that, in presence of a modular structure (Figures \ref{fig:Comp}
a,b and c)  in the network and weakly informative  features,
\texttt{CohsMix} algorithms always performs better than SMLV and HMRF
algorithms.

 Besides, it is
noticeable that performances of \texttt{CohsMix} increase with the number of
features and/or with the distance between mean vectors (Figure \ref{fig:Comp} b and
d). HMRF algorithm with Potts \emph{a priori}  use the neighborhood structure for smoothing the partition.  A vertex with all its neighbors of a given class has a
high probability to be assigned to this class but  HRMF does not take advantage
of the graph structure as fully as  \texttt{Cohsmix}. Our model is thus  mainly attractive and suitable for datasets  with an existing  graph structure.

When there is no graph structure at all and few informative features (Figure
\ref{fig:Comp} d ) the \texttt{Cohsmix} does not compare to HMRF or SMLV. The \texttt{Cohsmix} algorithm is more sensitive to the total absence of
graph structure than its competitor.

But in all other setup,  the quality of
partition estimation remains good  with different kind of models,
the \texttt{CohsMix} algorithm appears very attractive and suitable for structured graphs
with vertex features.  We shall see in the next section that this
algorithm also performs well on real web datasets.

\subsection{Real data}
Exhaustivity is an essential feature for information retrieval systems like Web search engines. However,
it appears that  ambiguous queries produce such a huge diversity in the responses that it is a real impediment to understanding. 
A common way to circumvent this situation is to organize search results into groups (clusters), one for each meaning of the query.
This concern has been in the focus of the information retrieval community \citep{hearst1996rch,zamir1998wdc}
since the early days of the Web. More recently, academic \citep{zeng2004lcw} and industrial \citep{bertin2002sta}
(\url{exalead.com} or \url{clusty.com}) attempts have made the clustering of search results a common feature
for a WWW user.


The main drawback of many Web page clustering methods is that they take into account only the
topical similarity between documents in the ranked list and they do not consider the  topology induced by hyperlinks.
But in competitive or controversial queries like "abortion" or "Scientology" such methods do not reveal community
information that is visible on the link topology : By affinity, authors tend to link to pages with similar
topics or points of view which create a thematic locality \citep{davison2000}. In addition, ambiguous queries
like "orange" or "jaguar" can also benefit from the link topology to produce more accurate separation of results.
Combination of topological and topical clustering methods is a proven strategy to build an
relevant system. One of the  most relevant previous work is suggested in \citet{he2002wdc}, which build a
Web page clustering system which accounts for  the hyperlinks structure of the Web, considering two Web pages
to be similar if they are in parent/child or sibling relations in the Web graph. A more general multi-agent
framework based on path between each pair of results has been proposed by \citet{bekkerman2006wpc}, but these
methods, not model-based, use various heuristics and fine tunings.

\paragraph{Datasets setup}
We use \url{exalead.com} search engine in our real data experiments. For each query, we retrieve the first 150
search results in order to build our graph and feature structures. Indeed, the web is a very sparse graph and
thematic subgraphs may amplify this  property creating unconnected components which inhibits the opportunity to
use classical graph clustering  algorithms directly on the observed adjacency matrix. In order to increase the
graph density, the probability  to have a link between two nodes, we propose to use  the site graph of \url{exalead.com}
basically based on  the concepts of \citet{raghavan2003rwg}. In this graph, nodes represent websites (a website contains
a set of pages) and edges represent hyperlinks between websites. Multiple links between to different website are collapsed
into a single link. Intra-domain links are taken into account if hostnames/websites are not similar. This site graph
is previously computed. This methodology is similar to the Exalead application called Constellations :
\url{constellations.labs.exalead.com}.  

Then, text features are extracted from the content of the web page
returned by the search engine.  The features are built using various text processing like
normalization, tokenization, entities detection, noun phrase detection
and related terms detection. Besides, we remove rare features which do not
appear more than twice.  Eventually each feature vector is approximately  of dimension $p=100$ and summarizes
all  text of  a returned page.

\begin{figure}[htb]
\begin{tabular}{c c}
\includegraphics[width=6cm]{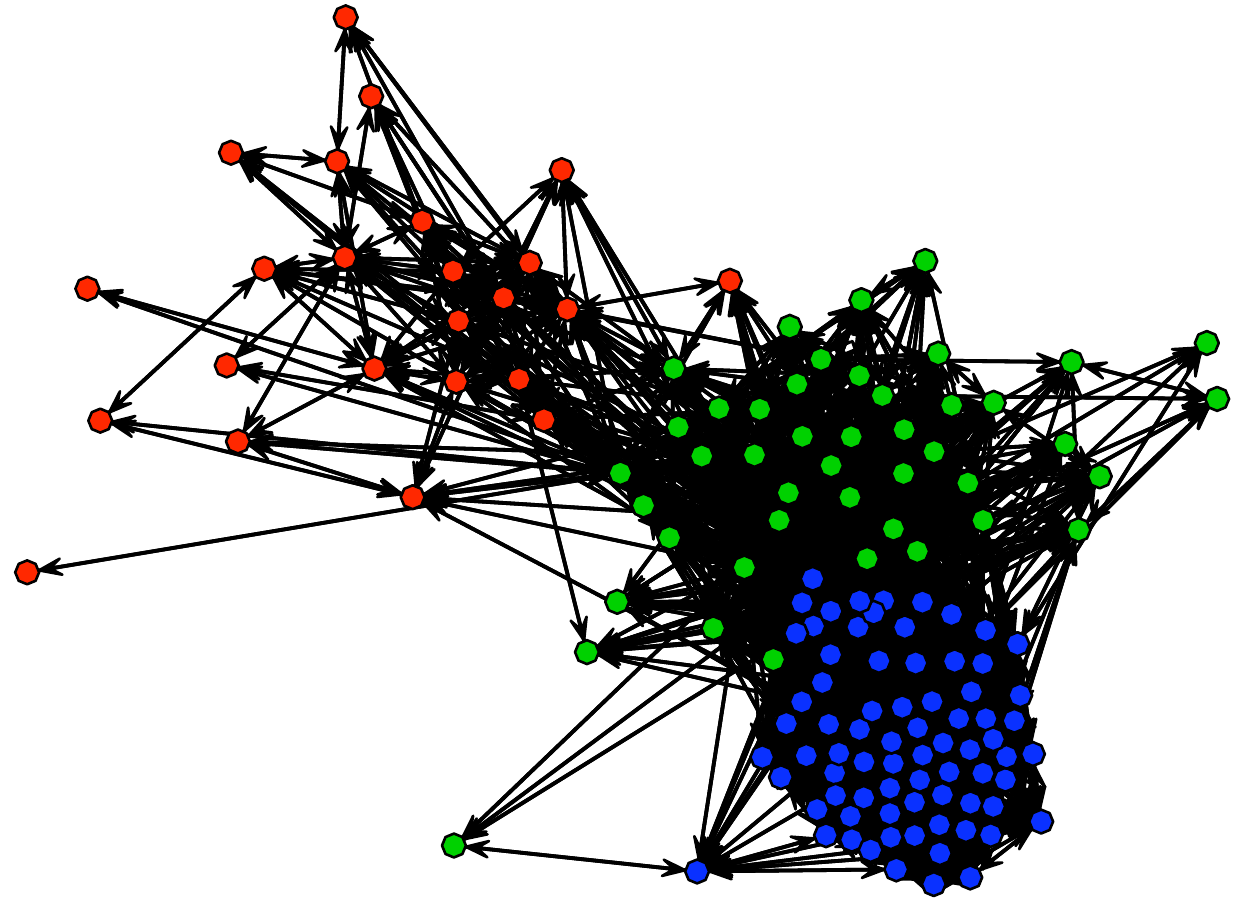} &  \includegraphics[width=6cm]{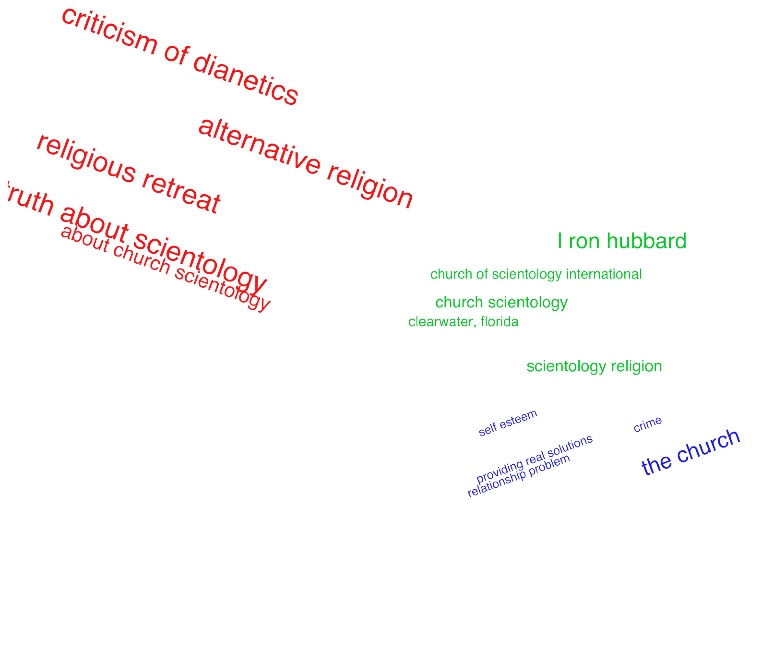} \\
\end{tabular}
\caption{Representation of the results of a  clustering of  the webpages returned by  the controversial query "Scientology" using  \texttt{CohsMix}. The graph structure is represented on the left  and on the  right are the main features. Colors indicate the \texttt{CohsMix} classification}
\label{fig:WebScientology}
\end{figure}

\paragraph{Algorithm results.}
We choose one ambiguous query ("jaguar")  and one controversial query ("Scientology") to illustrate
our algorithm behavior with real datasets. In Figure \ref{fig:WebScientology} associated to the query "Scientology",
we can observe a well structured graph  which fits our estimated latent partition with an optimal number of
classes $Q=3$. Basically this partition yields the pro- and anti-Scientology clusters and identifies a gateway cluster
(composed for example by \url{http://en.wikipedia.org/wiki/Scientology}) bridging the pro and anti cluster. Then, we concentrate
our attention on the most representative text features of each class $q$. To succeed, we select the best occurrence of
term features in the different $\mu_q$. Once again (see \ref{fig:WebScientology}), we notice terms describing pro- like
("self esteem or "providing real solutions") and anti- like ("criticism of dianetics" or "truth about Scientology").
The interface class is composed by common terms describing the church of Scientology. Thus, in a web context,
the \texttt{CohsMix} algorithm is enable to named the different found partitions which a precious assistance to have
rapidly a global overview of the hidden structure.

\begin{figure}[htb]
\includegraphics[width=6cm]{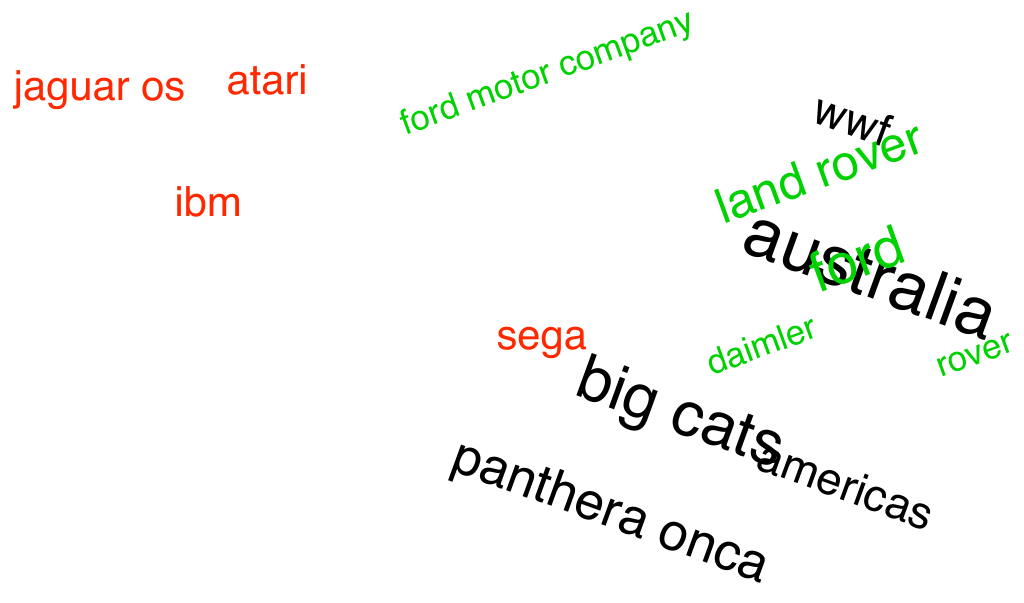}
\caption{Representation of the results of a  clustering of  the webpages returned by  the ambiguous query "jaguar".}
\label{fig:WebJaguar}
\end{figure}

The results of the processing of the  ambiguous query "jaguar" is represented in Figure \ref{fig:WebJaguar}. \texttt{CohsMix} clearly identifies three contexts: computer, animal and car model related web pages.

The above results illustrate that our algorithm \texttt{CohsMix} seems well adapted to detect ambiguous or controversial queries
of WWW search engine users.


%% file: conclusion.tex
\section{Conclusion}
This paper has proposed an algorithm for clustering dataset whose modelisation could be a graph structure
embedding vertex features. Characterizing each vertex both by a vector of features as well as by its
connectivity, \texttt{CohsMix} algorithm, based on a variational approach of EM, uses both elements to cluster
the data and estimate the model  parameters. When analyzing simulation and comparison results, our algorithm appears
very attractive and competitive for various kind of models.  We have tested \texttt{CohsMix} algorithm to
cluster web search results based on hypertextuality and content and we demonstrate good relevance of this model
approach. We find that our algorithm successively exploits whatever information is found both in the connectivity
pattern  and in the features. In the short-term, we plan to investigate how to focus an a type of information, graph or features, when
it gets the upper hand.